\documentclass[doublecol]{epl2}

\usepackage[english]{babel}
\usepackage{amsfonts}
\usepackage{amsmath}
\usepackage{graphicx}
\usepackage[caption=false]{subfig}

\usepackage{soul}
\usepackage{color}

\title{Geometric phase corrections on a moving particle in front of a dielectric mirror }

\author{Fernando C. Lombardo and Paula I. Villar}

\institute{                    
	Departamento de F\'{\i}sica Juan Jos\'e Giambiagi 
	 and IFIBA CONICET-UBA,
	 Facultad de Ciencias Exactas y Naturales, Ciudad Universitaria, 
	 Pabell\'on 1, 1428 Buenos Aires, Argentina 
	\footnotesize Electronic Address: {\tt lombardo@df.uba.ar}}

\abstract{We consider an atom (represented by a two-level system) moving in front of a dielectric plate, and study how traces of dissipation and decoherence (both effects induced by vacuum field fluctuations) can be found in the corrections to the unitary geometric phase accumulated by the atom.  We consider the particle to follow a classical, macroscopically-fixed trajectory and integrate over the vacuum field and the microscopic degrees of freedom of both the plate and the particle in order to calculate friction effects.  We compute analytically and numerically the non-unitary geometric phase for the moving qubit under the presence of the quantum vacuum field and the dielectric mirror.   We find a velocity dependence in the correction to the unitary geometric phase due to quantum frictional effects. We also show in which cases decoherence effects could, in principle, be controlled in order to perform a measurement of the geometric phase using standard procedures as Ramsey-like interferometry.}

\begin{document}

\maketitle

\section{Introduction}

A system can retain the information of its motion when it undergoes a cyclic evolution in the form of a
geometric phase, which was first put forward by Pancharatman in optics \cite{Pancharatnam} and later studied explicitly by Berry in
a general quantal system \cite{Berry}.  Since the work of Berry, the notion of geometric phases has been shown to have important consequences for quantum systems. Berry demonstrated that quantum systems could acquire phases that are geometric in nature. He showed that, besides the usual dynamical phase, an additional phase related to the geometry of the space state is generated during an adiabatic evolution. Since then, great progress has been achieved in this field. 
Due to its global properties, the geometric phase is propitious to  construct  fault tolerant  quantum  gates. In this
line of work, many physical systems  have  been  investigated  to
realise  geometric  quantum  computation,  such  as  NMR  (Nuclear
Magnetic Resonance) \cite{NMR}, Josephson  junction \cite{JJ},  Ion  trap \cite{IT} and
semiconductor  quantum  dots \cite{QD}. The quantum computation scheme for the 
geometric phase has been proposed based on the Abelian  or
non-Abelian geometric concepts, and the geometric phase has been shown
to be robust against faults in the presence of some kind of
external noise due to the geometric nature of Berry phase \cite{refs1, refs2, refs3}. It was
therefore seen that interactions play an important role in the
realisation of some specific operations. As the gates operate slowly compared to the 
dynamical time scale, they become vulnerable to open system effects and parameters fluctuations
that may lead to a loss of coherence. Consequently, study of the
geometric phase was soon extended to open quantum systems. Following this idea,
 many authors have analysed the correction to the unitary geometric phase under the influence of 
an external environment using different
approaches (see \cite{Tong, pra, nos, pau} and references therein). In this case, the evolution of an open quantum system is eventually plagued by non unitary features like decoherence and dissipation. Decoherence, in particular, is a quantum effect whereby the system loses its ability to exhibit coherent behaviour.

On the other hand, quantum fluctuations present in the vacuum are \mbox{responsible} for non-classical effects that can be experimentally detected \cite{libros} and give rise to numerous fascinating physical effects, in particular on sub-micrometer scales. Some of these phenomena have been extensively studied and carefully measured, thus demonstrating their relevance for both fundamental physics and future technologies \cite{Decca, IntraviaNat}. Over the past few years, an increasing attention has been paid to the Casimir forces and moving atoms \cite{scheel} and the interaction between a particle and a (perfect or imperfect) mirror \cite{others,dalvit,milonni,intravaia,pieplow,behuninhu,impens}, and also there have been works explaining how the non-additive vacuum phases may arise from the dynamical atomic motion \cite{impens2}. In this framework, it is of great interest to
calculate the frictional force exerted over the particle by the surface, mediated by the vacuum field fluctuations. As in the case of
the quantum friction between two plates \cite{Pendry97,debate,vp2007}, there is still no total agreement about the nature of this frictional
force. However, frictional and normal (Casimir) forces are not the only effects of the vacuum quantum fluctuations. These 
fluctuations can behave as an environment for a given quantum system, and due to this interaction, some traces of the quantumness of the  system can be destroyed via decoherence and consequently, a degradation of pure states into mixtures takes place. In the particular case of the vacuum field, it can not be switched off. Therefore,  any particle (whether charged or
with non-vanishing dipole moment) will unavoidably interact with the electromagnetic field fluctuations. The effects of the electromagnetic
field over the coherence of the quantum state of a particle, and the way in which this effect is modified by the presence of a conducting plate, may be studied by means of interference experiments \cite{pazmazzi,maianeto}.  Recently, in Ref.\cite{friction2}, the
decoherence process on the internal degree of freedom of a moving particle with constant velocity (parallel to a dielectric mirror) has been studied. 
 The loss of quantum coherence of the particle's dipolar moment becomes relevant in any interferometry experiment, where the depolarisation of the atom could be macroscopically observed by means of the Ramsey fringes \cite{ramsey,rabi}. 

In this framework, we propose to track evidence of \mbox{vacuum} fluctuations on the geometric phase acquired by a neutral particle moving in front of an imperfect mirror. By measuring the interference pattern of the particle, it could be possible to find a dependence of the correction to the unitary geometric phase upon the velocity of the particle. The pattern obtained in this model can be an indirect prove of the existence of a quantum frictional force. We shall consider a neutral particle coupled to a vacuum field, which is also in contact with a dielectric plate.
The particle's  trajectory will be, along this paper, kept as an externally-fixed variable.  We shall consider a toy model to analyze the 
plausibility of this novel idea.  In our model, the particle will move at a constant velocity $v$  (in units of $c = 1$, $v$ is dimensionless), as is the most popular scenario in the literature \cite{others, dalvit}. As we are interested in the dynamics of the internal degree of freedom of the particle,  we will consider the neutral particle as a two-level quantum system (a qubit, as in many models used to represent a real atom),  coupled to the vacuum field. We will also use a simple model for the microscopic degrees of freedom of the mirror, as we have done in a previous work \cite{friction}: a set of uncoupled harmonic oscillators, each of them also interacting locally with the vacuum field.  Despite this freedom from complexity, the model admits the calculation of some relevant quantities without much further assumptions. In order to consider how the relative motion between the particle and the plate affects the geometric phase acquired 
we shall follow the procedure presented in previous works \cite{friction, friction2}.

%%%%%%%%%%%%%%%%%%%%%%%%%%%%%%%%%%%%%%%%%%%%%%%%%%%%%%%%%%%%%%%%%%%%%%%
%%%%%%%%%%%%%%%%%%%%%%%%%%%%%%%%%%%%%%%%%%%%%%%%%%%%%%%%%%%%%%%%%%%%%%%
%%%%%%%%%%%%%%%%%%%%%%%%%%%%%%%%%%%%%%%%%%%%%%%%%%%%%%%%%%%%%%%%%%%%%%%
%%%%%%%%%%%%%%%%%%%%%%%%%%%%%%%%%%%%%%%%%%%%%%%%%%%%%%%%%%%%%%%%%%%%%%%
\section{Dissipative quantum friction}
\label{sec:inout}

\begin{figure}[h]
\centering
\includegraphics[scale=0.5]{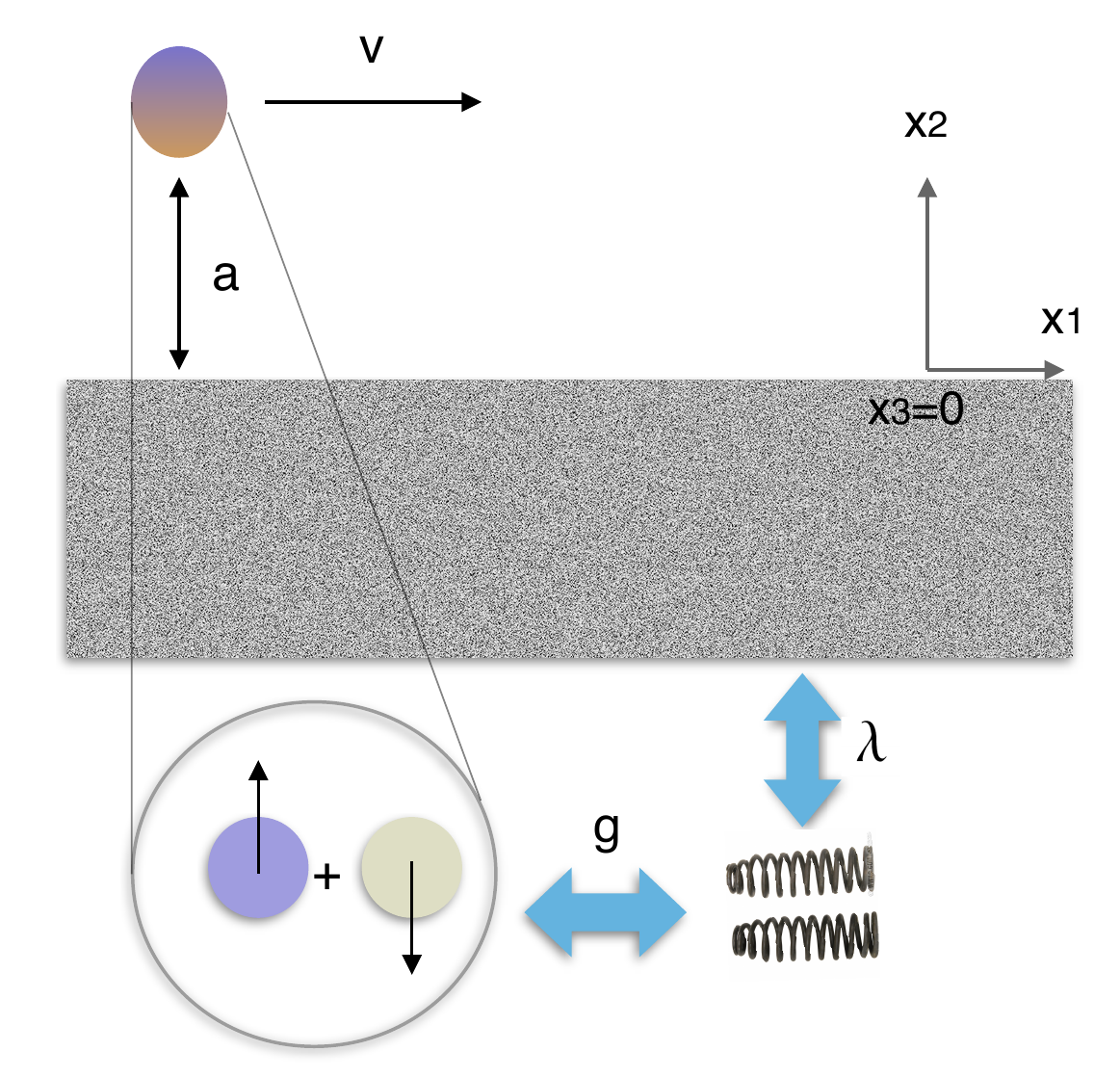}
\caption{ We present a simple diagram of the system under consideration. The vacuum field is a massless scalar field $\phi(x)$ and the internal degrees of freedom of the plate are $\psi(x)$. The internal degrees of freedom of the particle will be considered as a two-level system $\sigma_z$.}
\label{esquema}
\end{figure}

In the current Section, we shall assume the vacuum field to be a massless scalar field $\phi(x)$, interacting with both the 
particle and the internal degrees of freedom of the plate which are represented by $\psi(x)$ \cite{friction}.  The particle moves in a macroscopic, externally-fixed, uni-dimensional trajectory parallel to the plate, schematized as in Fig. \ref{esquema}. The distance $a$ between the particle and the plate is kept constant by an external source. 
The particle also has an internal degree of freedom that we shall call $\sigma_z$ in order to model a two-level system.  

We may write the classical action for the system as:
\begin{eqnarray}
S[\phi,\psi,\sigma_z] &=&S_0^{\text{vac}}[\phi]+S_0^{\text{pl}}[\psi]+S_0^{\text{at}}[\sigma_z]+S_{\text{int}}^{\text{vac-pl}}[\phi,\psi] \nonumber \\
&+&S_{\text{int}}^{\text{at-vac}}[\phi,\sigma_z] \, ,
\end{eqnarray}
where the action for the free vacuum field, is given by
$S_0^{\text{vac}}[\phi]=-\frac{1}{2} \int dx \phi(x) \left[ \partial_\mu \partial^\mu -i \epsilon \right] \phi(x) $.
After integrating out the degrees of freedom of the plate, we get an effective interaction potential $V(x,x')$ for the vacuum field, similarly to what has been done  in \cite{Fosco2011}. This procedure results in a new action defined as
$S[\sigma_z,\phi]=S_{\text{eff}}[\phi]+S_0^{\text{at}}[\sigma_z]+S_{\text{int}}^{\text{at-vac}}[\sigma_z,\phi] $,
with
$S_{\text{eff}}[\phi]=S_0[\phi]+\int dx dx' \phi(x) V(x,x') \phi(x') $.

As it has been considered previously in the Literature \cite{friction2}, the internal degree of freedom of the particle 
interacts with the vacuum field through a given current $j(x)$. The interaction term can then be written as 
$S_{\text{int}}^{\text{at-vac}}[\sigma_z,\phi]=i \int dx \phi(x) j(x) $. 
In \cite{friction2} it has been derived the in-out effective action for the particle, which is the action obtained after functionally integrating over the vacuum field and over the internal degrees of freedom of the dielectric plate (polarisation degrees of freedom) $\psi(x)$.  The reason to evaluate the in-out effective action is that this is related to the vacuum persistence amplitude, and  the presence of an imaginary part signals the excitation of internal degrees of freedom on the mirror. Since this is due to the constant-velocity motion of the 
particle, it reflects the existence of non-contact friction \cite{friction2,friction,Fosco2011}.

We shall now consider the internal degrees of freedom of the plate to be an infinite set of uncoupled harmonic oscillators of frequency $\Omega$  
(the set of harmonic oscillators is characterized by a spectral density with one predominant phononic mode). Each of these oscillators are interacting locally in position with the vacuum field through a coupling constant $\lambda$. The internal degree of freedom of the particle is a two-level system $\sigma_z$, also interacting linearly and locally with the vacuum field through a coupling constant $g$. In the case studied in Ref. \cite{friction2} the internal  degree of freedom of the moving particle has been considered as a harmonic oscillator with natural frequency $\omega_0$. In that case,  the imaginary part of the effective action  (up to second order in the coupling constants) is given by the following expression:

\begin{equation}
\text{Im}\Gamma_I \approx  \frac{T v \pi \lambda^2 \gamma_0}{32 \tilde{\Omega}\tilde{\omega}_0} 
\frac{e^{-\frac{2}{v}\sqrt{ (\tilde{\omega}_0 + \tilde{\Omega})^2 - v^2 \tilde{\Omega}^2}}}{ (\tilde{\omega}_0 + \tilde{\Omega})^2 - v^2 \tilde\Omega^2},\label{parteim}
\end{equation} 
where $T$ is total the time of flight of the particle; $\tilde{\Omega}=\Omega a$ and $\tilde{ \omega}_0 = \omega_0 a$ are the dimensionless frequencies (as we have defined above, $a$ is the distance between the particle and the plate). We have also  set the dissipative constant $\gamma_0 \equiv g^2$. As we mentioned above, this imaginary part of the effective action implies the excitation of internal degrees of freedom on the mirror which in turn impacts on the particle through the vacuum field \cite{Fosco2007}.  Due to the exponential in Eq.(\ref{parteim}),   dissipative effects are strongly suppressed as $v\rightarrow 0$.  This exponential vanishing of the dissipation effects has already been found, using different approaches, in previous works \cite{others,dalvit}. It is important to note that the coupling constant $g$ is the analogue to the electric dipole moment $d$ appearing in more realistic models, since it accounts for the interaction between the particle's polarisability and the electromagnetic (vacuum) field. In this sense, the results presented here correspond to the $d^2$ contribution to the friction. Lastly, let us recall that the $\lambda^2$ accounts for the interaction between the internal degrees of freedom of the mirror which completes the composite environment for the moving particle \cite{friction2}. It is worthy to stress that, even though ours is a toy model for the realistic problem in which the electromagnetic field shoud be considered, in Ref.\cite{graphene} the Casimir friction phenomenon in a system consisting of two flat, infinite, and parallel graphene sheets, which are coupled to the vacuum electromagnetic field has been considered. In fact, the transverse contribution to the 
imaginary part of the effective action in \cite{graphene} is qualitatively (as a function of $v$) similar to the one shown in Eq. (\ref{parteim}). 

The in-out effective action cannot be applied in a straightforward way to the derivation of the equations of motion, since they would become neither real nor causal. As is well known, in order to get the correct effective equations of motion and fluctuation effects, one should compute the in-in, Schwinger-Keldysh, or closed time path effective action (CTPEA) \cite{ctp}, which also has information on the stochastic dynamics, like decoherence 
and dissipative effects in a non-equilibrium scenario. By using the expression in Eq.(\ref{parteim}), we can evaluate the 
decoherence factor induced by the composite environment over 
a two-level system.  Herein, we shall consider the lowest energy labels of such 
oscillator in order to simulate the behaviour of a two-level system (we consider that the energy gap $\Delta$ between the 
excited and ground states of the two-level system is set as $\Delta \sim \omega_0$). Then, we can evaluate 
the imaginary part of the {\it influence action} \cite{ctp} for the two-level system, in the non-resonant case $\Omega \gg \omega_0$, 
from Eq. (79) in Ref. \cite{friction2}. The result is given by

\begin{equation}
\text{Im} S^{\text{IF}}  \approx \frac{ \gamma_0 T }{2}\left( 1 + \frac{ 2}{3} v^2 
+\frac{\lambda^2}{\tilde{\Omega}^3} v \frac{ e^{-\frac{2a\Omega}{v}\sqrt{1-v^2}}}{1-v^2}\right)
\label{noiseterm},
 \end{equation} here, $\gamma_0$ is dimensionless. 

It is important to note that the non-resonant case is not  the most 
decoherent case. As it has been shown in Ref. \cite{friction2}, the resonant case (for the harmonic oscillator internal degree of freedom) is 
the more effective case inducing loss of quantum coherence. Nevertheless, we use the non-resonant case in order 
to obtain an analytic expression for the decoherence factor and, consequently, for the corrections on the geometric 
phase. If one is able to find a velocity $v$ dependence in the correction to the phase, it would be an indication of the effect 
of quantum friction.  

Following standard procedures \cite{fer1}, it is possible to estimate decoherence time using Eq.(\ref{noiseterm}) when 
$\text{Im} S^{\text{IF}}(t = t_D)  \approx 1$ after evaluating in classical configurations. We shall evaluate decoherence 
factor using this procedure. The induced decoherence will modify the atom evolution in general and particularly, it will produce corrections on the unitary geometric phase of the atom states. These corrections will have two different sources:  a correction raised by the interaction with 
the  vacuum field and another one rooted in the interaction with the dielectric plate. These effects will 
be presented in the following section.

%%%%%%%%%%%%%%%%%%%%%%%%%%%%%%%%%%%%%%%%%%%%%%%%%%%%%%%%%%%%%%%%%%%%%%%
%%%%%%%%%%%%%%%%%%%%%%%%%%%%%%%%%%%%%%%%%%%%%%%%%%%%%%%%%%%%%%%%%%%%%%%
%%%%%%%%%%%%%%%%%%%%%%%%%%%%%%%%%%%%%%%%%%%%%%%%%%%%%%%%%%%%%%%%%%%%%%%
%%%%%%%%%%%%%%%%%%%%%%%%%%%%%%%%%%%%%%%%%%%%%%%%%%%%%%%%%%%%%%%%%%%%%%%
\section{Non-unitary geometric phase}
\label{nonunitaryGP}

We shall consider that the main qubit system (internal degree of freedom of the moving particle), can be represented by a bare Hamiltonian of the form $H_{\rm sys} = \Delta \sigma_z$, which simply represents a cyclic evolution with period $\tau=2\pi/\Delta$ if isolated; and we shall consider the effect of decoherence over this qubit. For simplicity, we are only considering a dephasing spin--bath interaction,
neglecting relaxation effects and limiting the relevance of the initial state (see discussion below).
We take a product initial state for the spin-bath system as
$\rho(0)=\vert{\varphi_0}\rangle \langle{\varphi_0}\vert\otimes\vert{\varepsilon(0)}\rangle \langle{\varepsilon(0)}\vert
$, where $\vert \varphi_0\rangle=\cos(\theta/2)\vert 0\rangle +\sin(\theta/2)\vert 1\rangle $ and $\vert \varepsilon(0)\rangle $ is a general
initial state of the composite bath. 
To compute the global phase gained during the evolution, one can use the Pancharatnam's definition \cite{Pancharatnam}, which has a gauge dependent part (i.e a {\textit {dynamical}} phase $\Phi_d=\pi \cos\theta$ and a gauge independent part, commonly known as {\textit{geometric}} phase (GP) $\Phi_g=\pi(1+  \cos\theta)$.

It is commonly known that, when coupled to a bath, the reduced density matrix for the particle system 
satisfies a master equation where non-unitary effects are included through noise and dissipation coeffients.  
For simplicity, we will assume that the model considered describes a purely decoherent mechanism. 
Therefore, the coupling to the bath affects the system such that its
reduced density matrix at a time $t$  is \cite{pra}

\begin{eqnarray} 
\rho_{\rm r}(t) &=& \cos^2(\theta_t) \vert {0}{0}\rangle + \sin^2(\theta_t)
\vert{1}{1}\rangle + \nonumber \\  && 
 \sin(\theta_t) \cos(\theta_t) e^{-i\Delta t}   \vert{0}{1}\rangle + \sin(\theta_t) \cos(\theta_t) 
e^{i\Delta t} \vert{1}{0}\rangle,  \nonumber \label{rhor}
\end{eqnarray}
where we have defined 
\begin{eqnarray}
\sin(\theta_t)&=&\frac{2 (\varepsilon_{+}-\cos^2(\frac{\theta}{2}))}{
\sqrt{|r(t)|^2 \sin^2(\theta)+4 (\varepsilon_{+}-\cos^2(\frac{\theta}{2}))^2}} \label{autoval0}\\
\cos(\theta_t)&=&\frac{|r(t)| \sin(\theta)}{\sqrt{|r(t)|^2 \sin^2(\theta)+4 (\varepsilon_{+}-\cos^2(\frac{\theta}{2}))^2}}, \label{autoval1}
\end{eqnarray} 
that encode the effect of the environment through the decoherence factor $r(t)$. Non-diagonal terms decay with $r$.
The eigenvalues of the above reduced density matrix
 are easily calculated, yielding:
\begin{equation}
\varepsilon_{\pm}(t) = \frac{1}{2} \pm \frac{1}{2}
\sqrt{\cos^2(\theta)
+ \vert r(t)\vert^2 \sin^2(\theta)}. \label{autoval2}
\end{equation}

The phase $\Phi $ acquired by the open system after a period $\tau$ is defined in the kinematical approach 
\cite{Tong} as, 
\begin{eqnarray}
\Phi = &\arg & \left[ \sum_{k}
\sqrt{\varepsilon_k(\tau) \varepsilon_k(0)} \langle \Psi_k(0)\vert \Psi_k(\tau)\rangle \right. \nonumber \\
&\times & \left. e^{-\int_0^\tau dt \langle \Psi_k(t)\vert \frac{\partial}{\partial t}\vert \Psi_k(t)\rangle
} \right], \label{gp}\end{eqnarray} 
where $ \vert \Psi_k(t)\rangle$ and $\varepsilon_k(\tau)$
are respectively the instantaneous eigenvectors and eigenvalues of $\rho_{\rm r}(t)$. Here,
$k$ refers to the two modes ($+$ and$-$) of the one qubit model we are dealing with. In order to estimate the geometric phase, 
we only need to consider the
eigenvector $|\Psi_{+}(t) \rangle$ since $\varepsilon_{-}(0)=0$. Therefore, in this case, the $+$ mode is the only contribution to the GP. By inserting Eqs.(\ref{autoval0}-\ref{autoval2}) into definition (\ref{gp}), one can straighfordwarly reach the 
final formula for the GP

\begin{equation}
\Phi =  \Delta  \int_0^\tau dt\cos^2(\theta_t).
\label{FinalGeometricPhase}
\end{equation}
The central result of Eq. (\ref{FinalGeometricPhase}) is the 
geometric phase, that reduces to the known results in the limit of an unitary evolution \cite{Tong}.
At this point, we are left with the definition of the decoherence factor $r(t)$ in order to proceed to the computation of the GP. It is possible to evaluate the decoherence factor from Eq.(\ref{noiseterm}) since $r(t) = \exp[- \text{Im} S^{\text{IF}}(t)]$ \cite{fer1}, 
where we will set dimensionless coupling constant to the dielectric plate ${\tilde \lambda}^2 = \lambda^2/\Omega^3$.

\begin{figure}[h]
\centering
\includegraphics[scale=0.6]{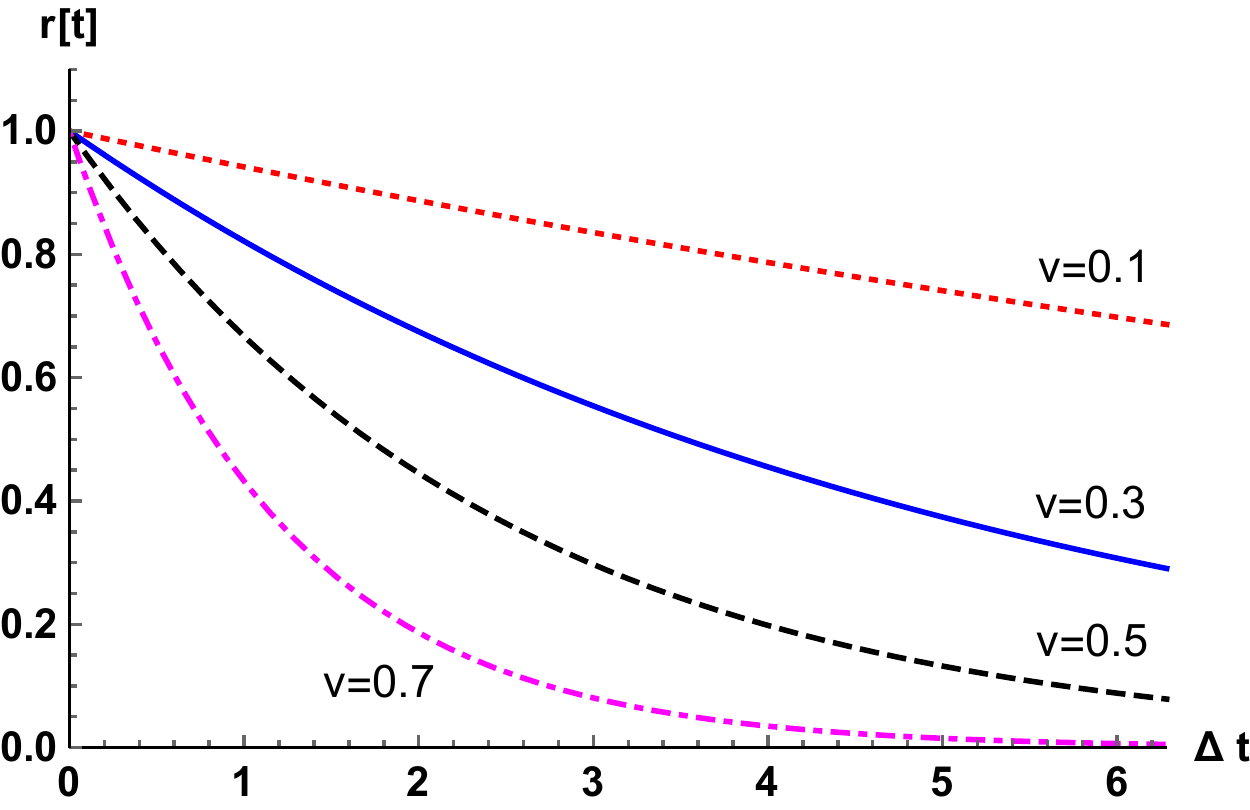}
\caption{\label{r(t)} Decoherence factor $r(t)$ for different values of the velocity $v$.
As we can see, the bigger the velocity of the two-state particle, the bigger the\mbox{ decoherence} rate. Parameters used: $\tilde \lambda=5$, $\gamma_0=0.05$ and $\tilde \Omega=0.03$.}
\end{figure}

In Fig.\ref{r(t)} we present the behaviour of the decoherence factor as a function of time (in units of $\Delta$) for different values of the velocity parameter. It is possible to note that as the particle completes one cycle of evolution, the decoherence is more destructive the more velocity the particle has. In two periods time, decoherence is strong enough in most cases, even for $v\ll 1$.

\begin{figure}[h]
\centering
\includegraphics[scale=0.6]{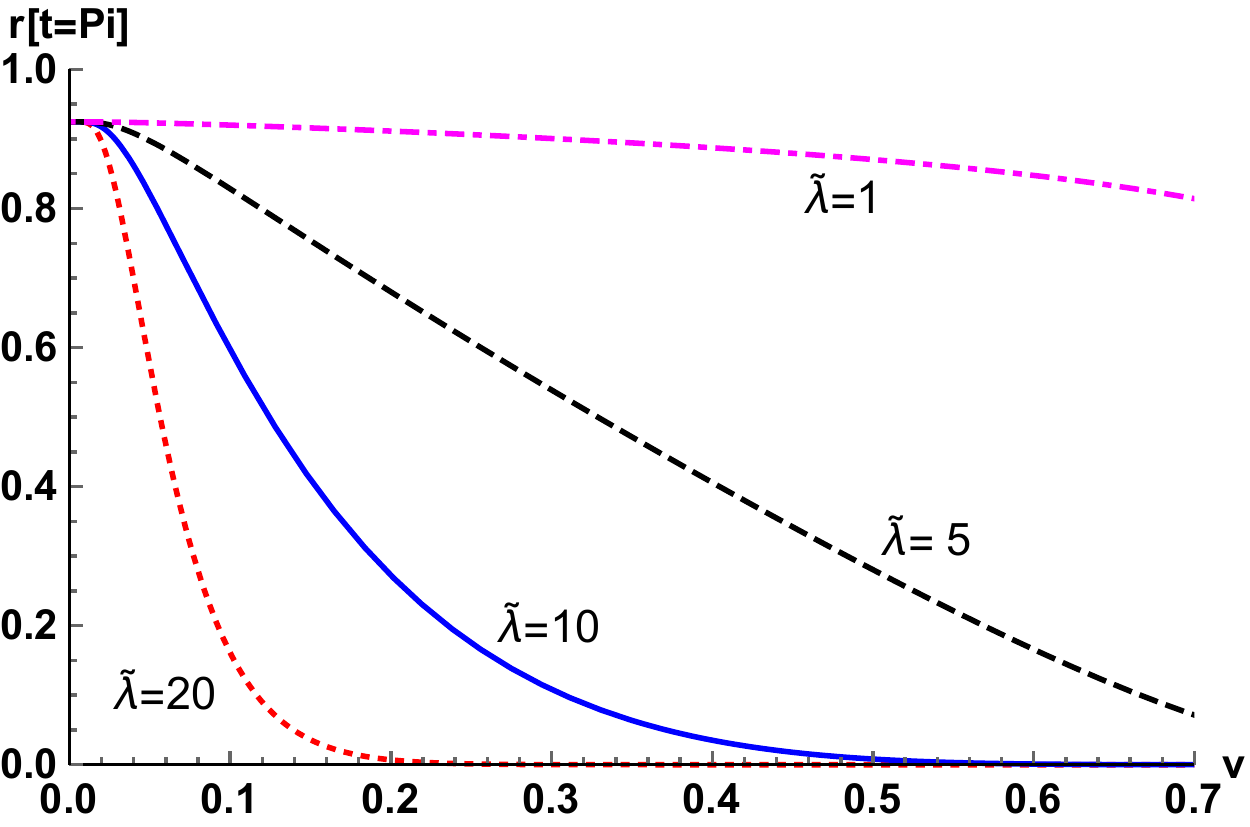}
\caption{\label{rtiempoPi} Decoherence factor for a time $\Delta t=\pi$ as function of the velocity $v$ for different values of ${\tilde \lambda}$. Parameters used: $\gamma_0=0.05$ and $\tilde \Omega=0.03$.}
\end{figure}

Besides the typical correction to the unitary phase due to the vacuum field (which is just proportional to the 
dissipative constant $\gamma_0$ and here there is an extra-dependence with $v^2$), there is also a term proportional to the quadratic power of the coupling between the vacuum field and 
the dielectric mirror (${\cal O}( \lambda^2)$). In this latter contribution there is also a dependence on the velocity of the atom ($v$). As expected, this contribution becomes less important when $v\rightarrow 0$ and grows for large values of $v$. 
In Fig.\ref{rtiempoPi} we show the decoherence factor for a fixed time $\Delta t=\pi$ for different values of the interaction coupling constant $\lambda$. We can see that even if this interaction is low, at half a period of isolated evolution, decoherence is non negligible even for small velocities of the particle. Then, we see that in this problem setting it is important to consider both features: the velocity the particle is travelling and  the time it takes to traverse, in addition to the parameters involved in the noise decoherence factor.

\begin{figure}[h]
\centering
\includegraphics[scale=0.6]{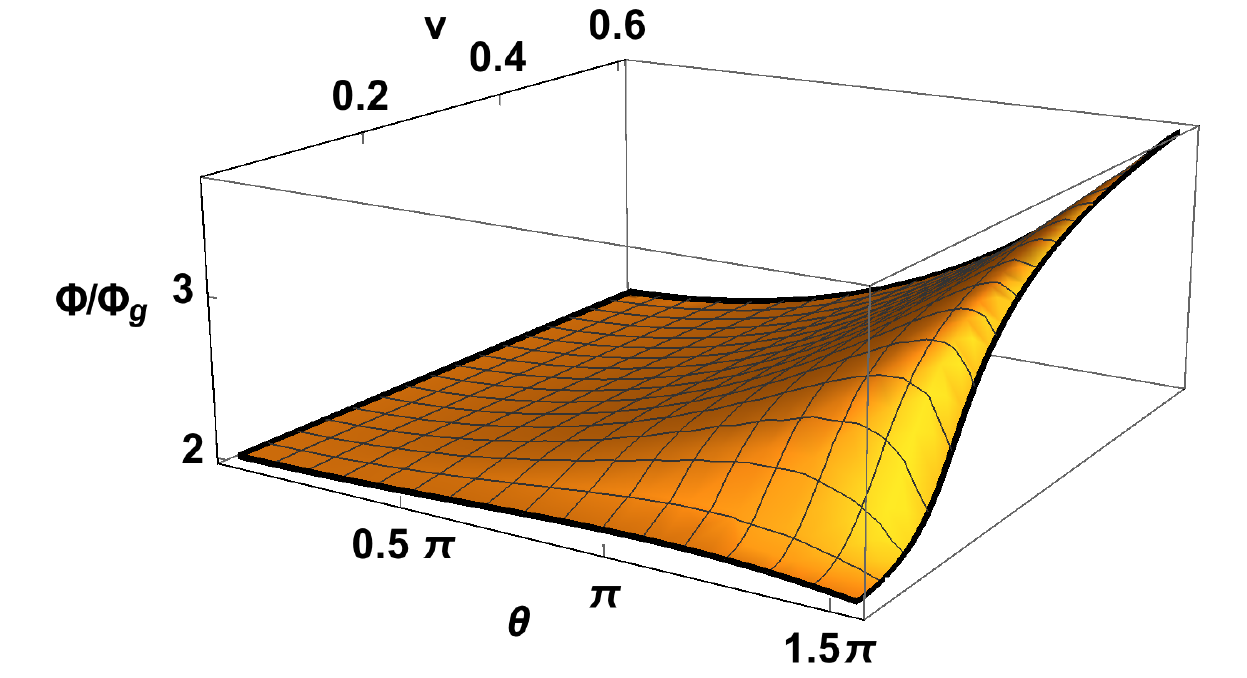}
\caption{ Geometric phase normalised by the unitary geometric phase $\phi_g=\pi(1+\cos\theta)$,  ($\Phi/\Phi_g$ )  as function of $\theta$ and $v$, considering a fixed time $\tau = 2\pi/\Delta$. Parameters used: $\tilde \lambda=15$, $\gamma_0=0.05$ and $\tilde \Omega=0.03$.} \label{3D} \end{figure}

In Fig.\ref{3D}  we plot the ratio between the total geometric phase from Eq.(\ref{FinalGeometricPhase}) and the unitary phase $\Phi_g=\pi(1+  \cos\theta)$, as a function of the initial angle $\theta$ and the tangential velocity $v$ for fixed parameters of $\tilde \lambda$, $\tilde \Omega$, $\gamma_0$, and time $\tau=2\pi/\Delta$ (period of the isolated evolution). Therein, we can see that for small values of the initial angle (i.e. a spin very similar to $|\uparrow \rangle$) and very low values of the velocity, the GP obtained for this system is very similar to the one obtained for an isolated quantum system (i.e. a spin $1/2$ particle evolving freely). The bigger difference between the open GP and the isolated one is seen for bigger angles and bigger values of $v$.

\begin{figure}[h]
\centering
\includegraphics[scale=0.6]{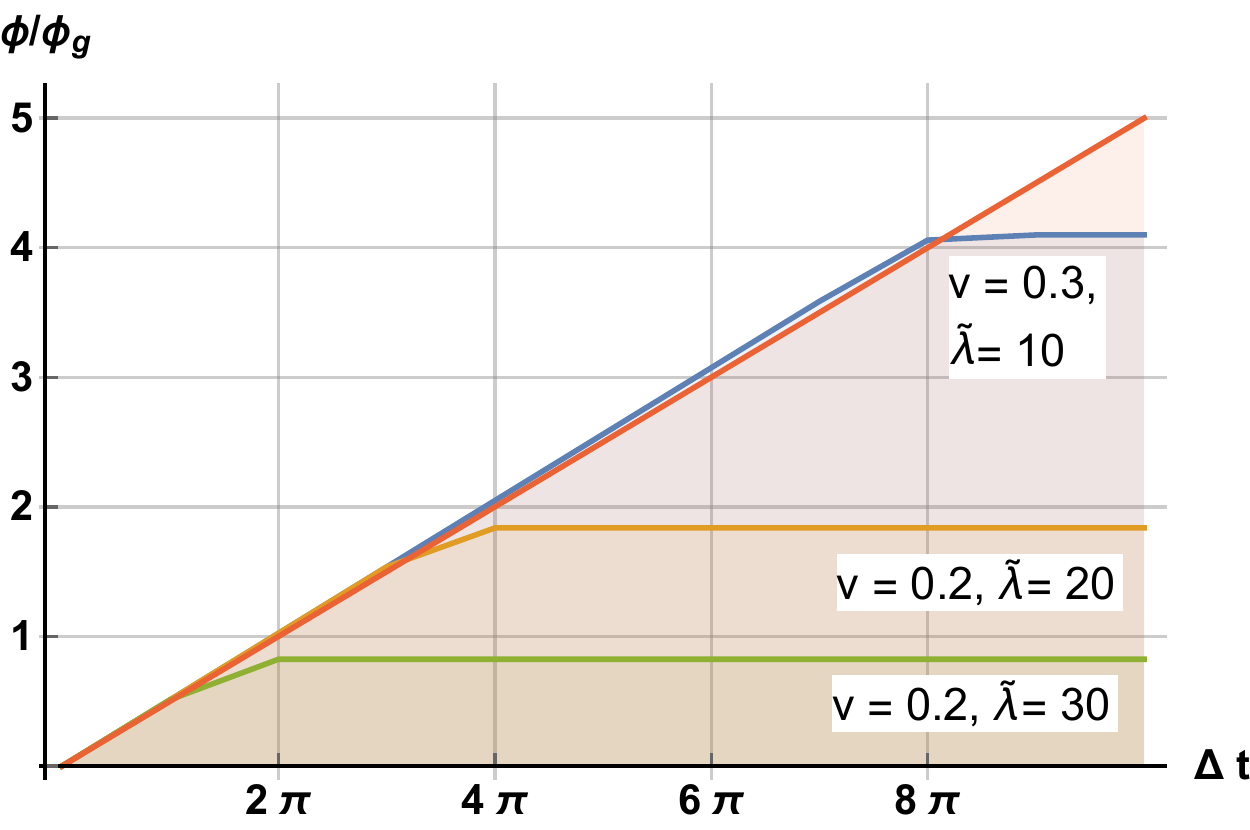}
\caption{\label{fase3Dt} Geometric phase normalised with the unitary phase $\phi_g= \pi(1+\cos\theta)$ (evaluated at one period of the isolated evolution $\tau=2\pi/\Delta$) as function 
of time, for different values of the velocity $v$ and the coupling constant $\tilde \lambda$. The straight (orange) line is the behaviour with time of the geometric phase when the system is isolated from the environment. The effect of the environment on the geometric phase 
can be clearly seen for bigger values of $\tilde \lambda$ and takes longer (more than a single period of the unitary evolution) for 
smaller couplings. Parameters used: $\theta = 0.1\pi$, $\gamma_0=0.1$, and $\tilde \Omega=0.03$.}
\label{niveles1}\end{figure}

The GP can be also analyzed as a function of time, for different values of the coupling constants $\tilde \lambda$ and $\gamma_0$ 
as well as the velocity $v$. In Fig.\ref{niveles1}  we plot the GP as function of time normalized with the unitary 
phase $\phi_g= \pi(1+\cos\theta)$ (evaluated at $\tau=2\pi/\Delta$) for different values of the velocity $v$ and the coupling constant $\tilde \lambda$. The straight (orange) line is the evolution with time of the GP when the system is isolated from the environment (evolves freely). The effect of the environment on the GP can be clearly seen for bigger values of $\tilde \lambda$ and takes longer (more than a single period of the unitary evolution $\tau$) for smaller couplings. 

\begin{figure}[h]
\centering
\includegraphics[scale=0.6]{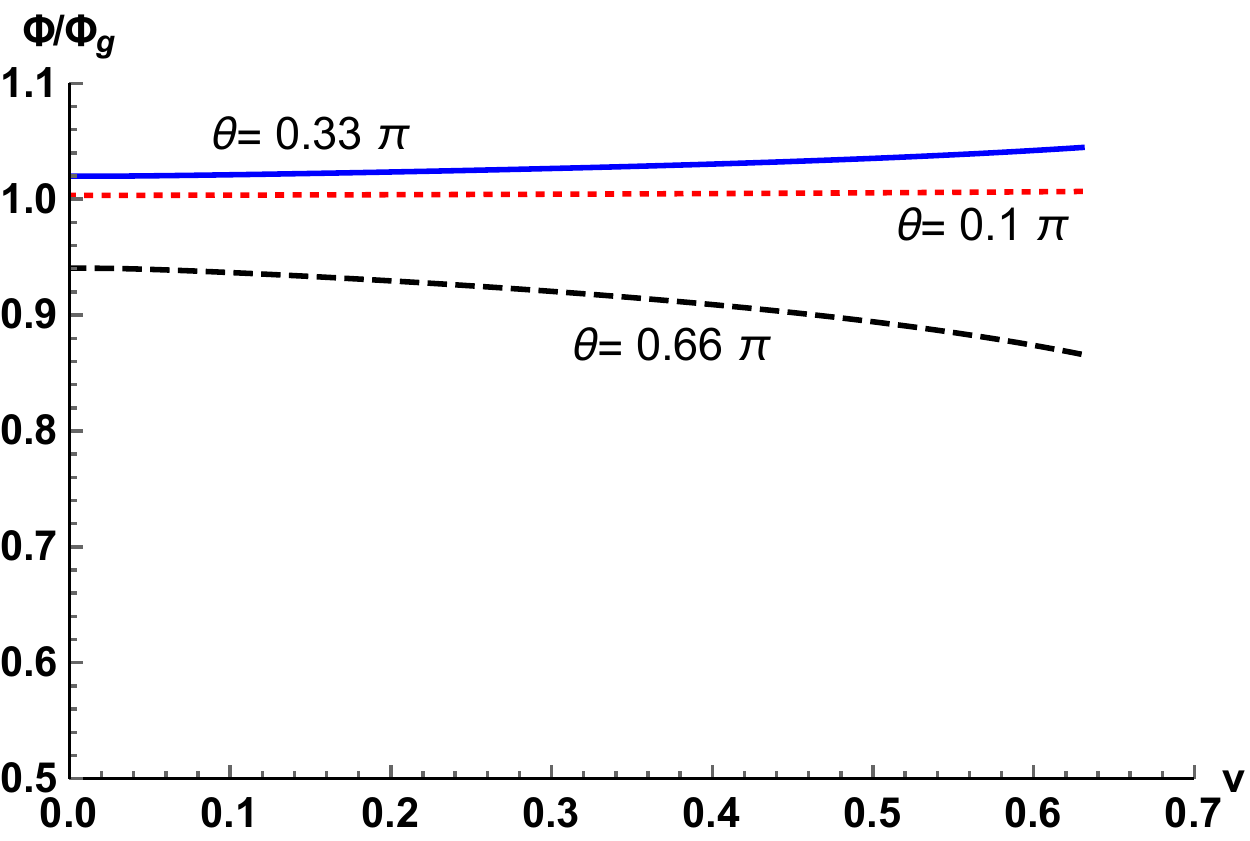}
\caption{\label{faselambda1} Normalised geometric phase $\Phi/\Phi_g$ for different initial angles $\theta$ as function of the velocity $v$ for a lower parameter $\tilde \lambda$. Parameters used:  $\tilde \lambda=1$, $\gamma_0=0.05$ and $\tilde \Omega=0.03$.}
\end{figure}

\begin{figure}[h]
\centering
\includegraphics[scale=0.6]{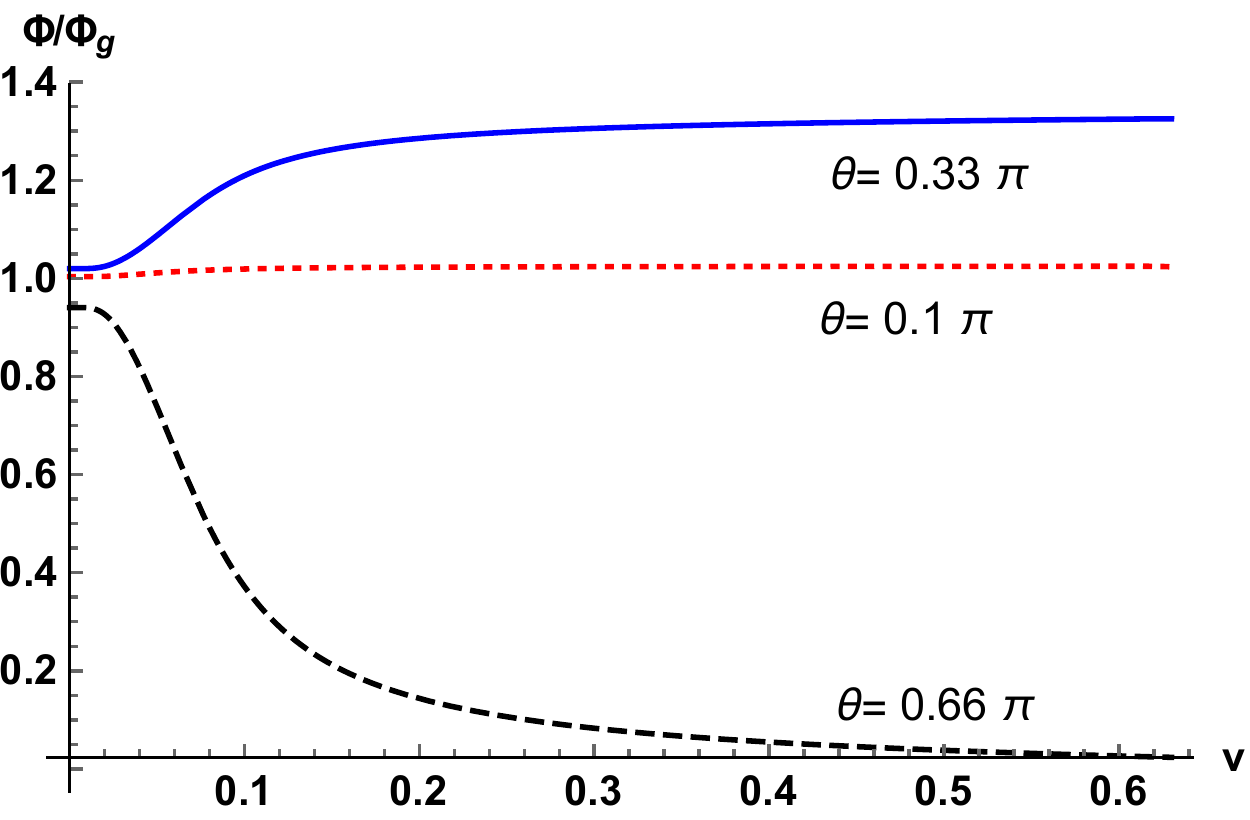}
\caption{\label{faselambda5} Normalised geometric phase $\Phi/\Phi_g$ for different initial angles $\theta$ as function of the velocity $v$ for a bigger parameter $\tilde \lambda$. Parameters used:  $\tilde \lambda=5$, $\gamma_0=0.5$ and $\tilde \Omega=0.03$.}
\end{figure}

In Figs. \ref{faselambda1} and \ref{faselambda5} we  show the dependence of the ratio between total and 
unitary geometric phase as a function of the 
tangential velocity of the particle for fixed initial angles $\theta$.  We can see that very small angles of the initial state of the particle, do not really suffer the difference between a lower or bigger value of the coupling constant $\lambda$. In those cases, what really matters is the coupling constant $\gamma_0$ and the velocity of the particle. All other angles are affected by the couplings constants $\gamma_0$ and $\lambda$, being more considerable when the velocity is greater.
Once more, we note that the correction is relevant for 
bigger values of $v$. It is important to remark that the mere presence of a velocity contribution to the phase, 
is an indication of the friction effect over the quantum degree of freedom of the atom. In this sense, 
the measurement of the geometric phase could, in principle, be an alternative way to find out quantum 
friction in a laboratory, even though the  velocities considered in experiments are still far away from 
a relativistic case with $v \rightarrow 1$.

Finally, we can perform a series expansion in $\gamma_0$ and $\lambda$ (up to first non-trivial orders) in Eq.(\ref{FinalGeometricPhase}) 
in order to obtain an analytical expression of the correction to the unitary geometric phase. The result for the geometric phase is then given by
\begin{eqnarray} 
\Phi_{\rm{approx}} &\approx & \pi (1 + \cos\theta) + \frac{\pi^2}{3} \gamma_0 \cos\theta \sin^2\theta 
\nonumber \\
&\times & \left[3 +2 v^2 + 2 v {\tilde\lambda}^2 (1 - v^2) e^{-\frac{2{\tilde \Omega}}{v}\sqrt{1-v^2}}\right].\label{perturbative}
\end{eqnarray} 

In the particular case in which the coupling between the atom and the dielectric plate is switched off, $\lambda = 0$, 
the correction to the unitary phase is given by $\delta\Phi \sim \pi^2 \gamma_0 (1+ 2/3 v^2)] \, \cos\theta \sin^2\theta$ which agrees with the  
result obtained for a two-level system coupled to an environment composed by an infinite set of harmonic 
oscillators at equilibrium with $T=0$ \cite{pra}. However, it is enhanced by the factor $1 + 2/3 v^2$. This situation  corresponds to the case where the atom is only coupled to the vacuum field. Up to the lowest perturbative order, the same result can be obtained in the limiting case of $v\rightarrow 0$. 

\begin{figure}[h]
\centering
\includegraphics[scale=0.6]{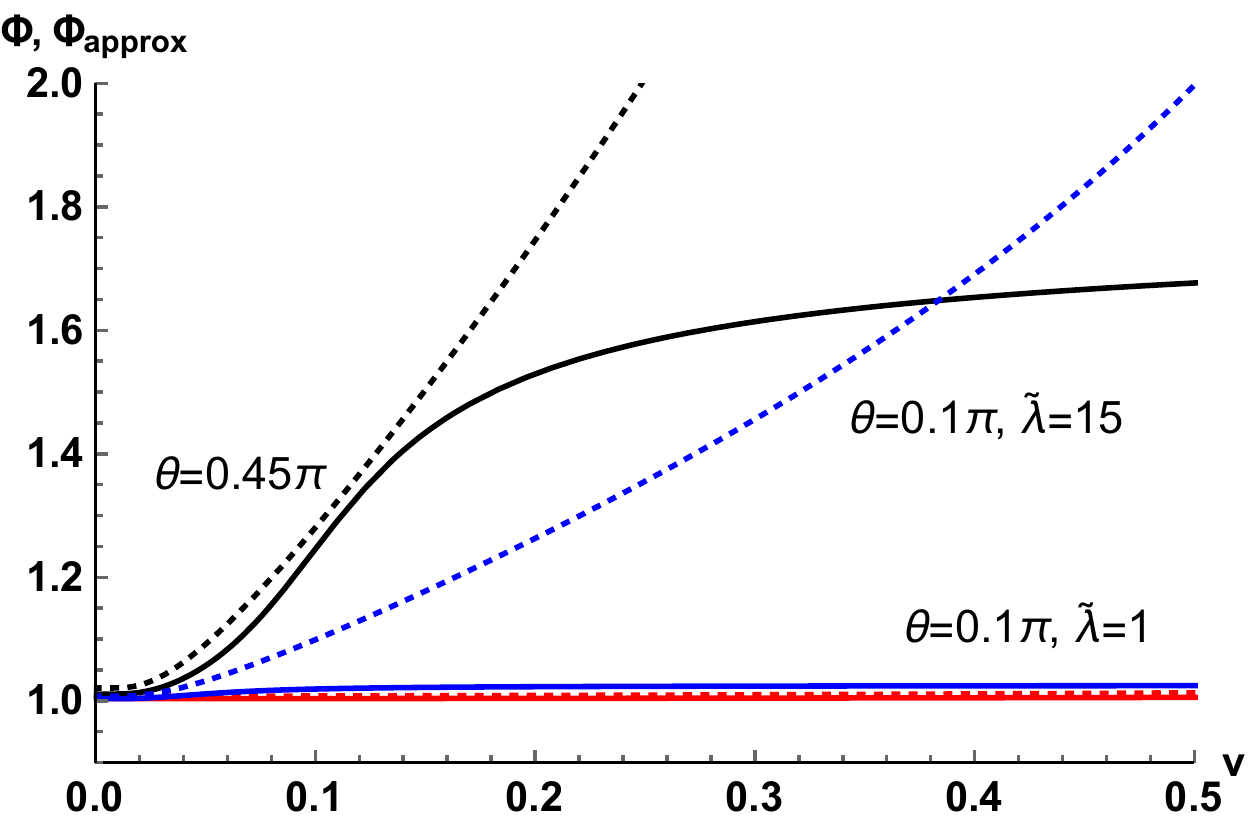}
\caption{\label{comparacion} Comparison between the geometric phase $\Phi$ (Eq.(\ref{FinalGeometricPhase})) (solid curves) and the analytical approximation $\Phi_{\rm approx}$ (dotted curves) obtained in Eq.(\ref{perturbative}) for different initial angles $\theta$. Parameters used:  $\tilde \lambda=5$, $\gamma_0=0.5$ and $\tilde \Omega=0.03$.}
\end{figure}

In Fig.\ref{comparacion} we plot the ratio between the geometric phase from Eq. (\ref{FinalGeometricPhase}) and 
the result in Eq.(\ref{perturbative})  as a function of the the velocity of the qubit $v$ for different values of the coupling constant $\tilde \lambda$ and initial angles $\theta$. Given a same value of the coupling constant $\gamma_0$, we can see that for a small angle $\theta=0.1 \pi$, there is still a noticeable difference between the behaviour of the GP for different values of $\tilde \lambda$.  In the case of $\theta=0.1 \pi$ and $\tilde \lambda=15$, we can even note that the approximate expression of the GP does not hold any longer.  For $\theta=0.45 \pi$ and $\tilde \lambda=5$, the approximate expression holds for small values of the velocity parameter $v$.  For a small angle and a low value of $\tilde \lambda$, the exact and approximate expression are very similar for all values of the velocity. 

\section{Conclusions}
\label{sec:conc}

We have considered a simple model to study the effects of quantum vacuum
fluctuations on a particle moving parallel to an imperfect mirror. In the model, the vacuum field is taken as a 
massless scalar field coupled to the microscopic degrees of freedom of the
mirror and the internal degree of freedom of the particle. The plate is formed by uncoupled
unidimensional harmonic oscillators, each of them interacting locally in position with the vacuum field. The
macroscopic trajectory of the particle is externally fixed, and its internal degree of freedom is considered 
as a two-level system, also coupled to the scalar field. 
Using previous results for the dissipative and decoherence effects reported in Refs.\cite{friction,friction2}, we have 
estimated the decoherence factor when the internal degree of freedom of the particle moving with parallel 
velocity $v$ is a two-state system. Once obtained an  expression for the 
decoherence factor, it was possible to calculate the corrections to the geometric phase acquired by
 the atom, induced by the interaction with the composite environment. 

In our analysis of the decoherence factor we have shown its functional dependence on different parameters involved in the model, such as: the coupling constant between the quantum system and the quantum field ($\gamma_0$), the coupling constant between the vacuum field and the imperfect mirror ($\lambda$) and the velocity of the quantum particle $v$.
We have shown that all these parameters contribute to a major decoherence rate in different ways.
Furthermore, we have computed the geometric phase acquired by the spin-$1/2$ in an open evolution. We have observed that the more interaction between the vacuum field and the plate, the more corrected results the phase acquires. This means that the phase of the quantum particle is different to the one the particle would have acquired if it had evolved freely. 
By measuring the correction to the unitary geometric phase, one can get an insight of the dependence of the phase on the parameters modified. In this way, we have seen that the bigger the velocity of the particle, the more correction to the phase.
It is also noticeable that the effect of noise is bigger for initial states near the equator of the Bloch sphere.

Finally, we have obtained an approximate analytical expression for the phase acquired (in a power expansion in the coupling constants) and compared this result to the exact geometric phase. In this case, we have seen that the expression gives an accurate result in the case of small values of the coupling constants (as expected), as well as small angles.

We expect that, in a Ramsey-like interference experiment, the parameters of our model could be chosen in a way that would maximize the decoherence effects. It is possible to choose its characteristic frequency close to the resonance with the plate. In addition, with a nonvanishing relative velocity, we expect decoherence effects to be observed by means of the attenuation of the contrast in the Ramsey fringes, after eliminating dynamical phase by spin-echo techniques.  By increasing the decoherence effect, the unitary geometric phase results in a major correction. 
In this way, as quantum friction has not been measured in labs yet, 
we expect that an indirect evidence could be obtained from measuring the environmental induced corrections to the 
geometric phase. The  dependence of the correction on the velocity $v$ would be an indirect way to measure quantum friction. 

\acknowledgments
We acknowledge support from CONICET, UBACyT, and ANPCyT (Argentina).

\end{document}